\documentclass[letterpaper]{article} 
\usepackage{aaai2027}
\usepackage[hyphens]{url}  
\usepackage{graphicx} 
\usepackage{natbib}  
\usepackage{caption} 
\usepackage{algorithm}
\usepackage{algorithmic}
\usepackage{amsfonts}
\usepackage{amsmath}
\usepackage{newfloat}
\usepackage{listings}
\DeclareCaptionStyle{ruled}{labelfont=normalfont,labelsep=colon,strut=off} 
\floatstyle{ruled}
\newfloat{listing}{tb}{lst}{}
\floatname{listing}{Listing}

\usepackage{booktabs}

\title{Turn-Based Structural Triggers: Structure-Conditioned Backdoors in Multi-Turn LLMs}
\author{
    Yiyang Lu\textsuperscript{\rm 1},
    Jinwen He\textsuperscript{\rm 1}\corresponding,
    Yue Zhao\textsuperscript{\rm 1}\corresponding,
    Kai Chen\textsuperscript{\rm 1},
    Ruigang Liang\textsuperscript{\rm 1},
    Yingjun Zhang\textsuperscript{\rm 2},
    Cheng Hong\textsuperscript{\rm 3}
}

\affiliations{
    \textsuperscript{\rm 1}Institute of Information Engineering, Chinese Academy of Sciences, China\\
    \textsuperscript{\rm 2}Institute of Software, Chinese Academy of Sciences, China\\
    \textsuperscript{\rm 3}Ant Group, China

}

\begin{document}

\maketitle

\begin{abstract}

Large Language Models (LLMs) are increasingly deployed as multi-turn assistants and customized through instruction tuning with project-specific training components. This practice creates a supply-chain risk when organizations reuse third-party fine-tuning frameworks, trainer extensions, or outsourced training code: an
adversary who subtly compromises the loss-computation component can inject malicious supervision during fine-tuning while leaving the stored training corpus, model architecture, and deployment interface unchanged. Existing LLM backdoors and defenses are largely prompt-centric, relying on lexical, syntactic, or semantic patterns in user inputs while overlooking structural signals in multi-turn conversations. We propose \textbf{Turn-based Structural Trigger (TST)}, a prompt-free backdoor that uses dialogue turn position as its activation condition. TST exploits structural cues implicitly encoded by chat templates and is implanted without modifying the stored dialogue corpus. Its trigger is automatically present once the conversation reaches the attacker-specified turn, making activation independent of downstream user inputs and resistant to prompt filtering, sanitization, and paraphrasing. In our primary setting, the model behaves normally during early interactions and activates the attacker-defined behavior only after the conversation reaches the designated structural condition. This delayed activation allows the backdoored model to pass prompt-centric checks, shallow multi-turn auditing, and standard utility evaluations while executing malicious behavior in deeper interactions. Across four open-source LLM families, TST achieves an average Attack Success Rate of 98.10\% on target turns and a Clean Rate of 99.96\% on non-target turns, while retaining 97.78\% of clean-model utility. It also generalizes across dialogue datasets, supports multiple malicious payloads and hybrid triggers, and remains effective under representative defenses. These results identify dialogue structure as an overlooked attack surface and motivate structure-aware auditing beyond prompt inspection.
\end{abstract}

\section{Introduction}

Large Language Models (LLMs)~\cite{touvron2023llamaopenefficientfoundation} are widely deployed as multi-turn assistants, where a \textit{user} and an \textit{assistant} exchange messages over several rounds~\cite{openai2024gpt4technicalreport}. In practice, organizations often fine-tune open-source models using standard frameworks with customized trainers or loss functions. This creates a supply-chain risk: an adversary who compromises the loss-computation code can implant hidden behaviors without modifying the stored dataset. The compromised code may keep the model benign during routine tests or the first several dialogue turns, but trigger unsolicited advertisements, harmful content, or degraded responses at one or more attacker-specified turns. As a result, the model can pass standard benchmarks and shallow dialogue evaluations while damaging service reliability and user trust after deployment.

\begin{figure*}[t]
    \centering
    \includegraphics[width=0.7\textwidth]{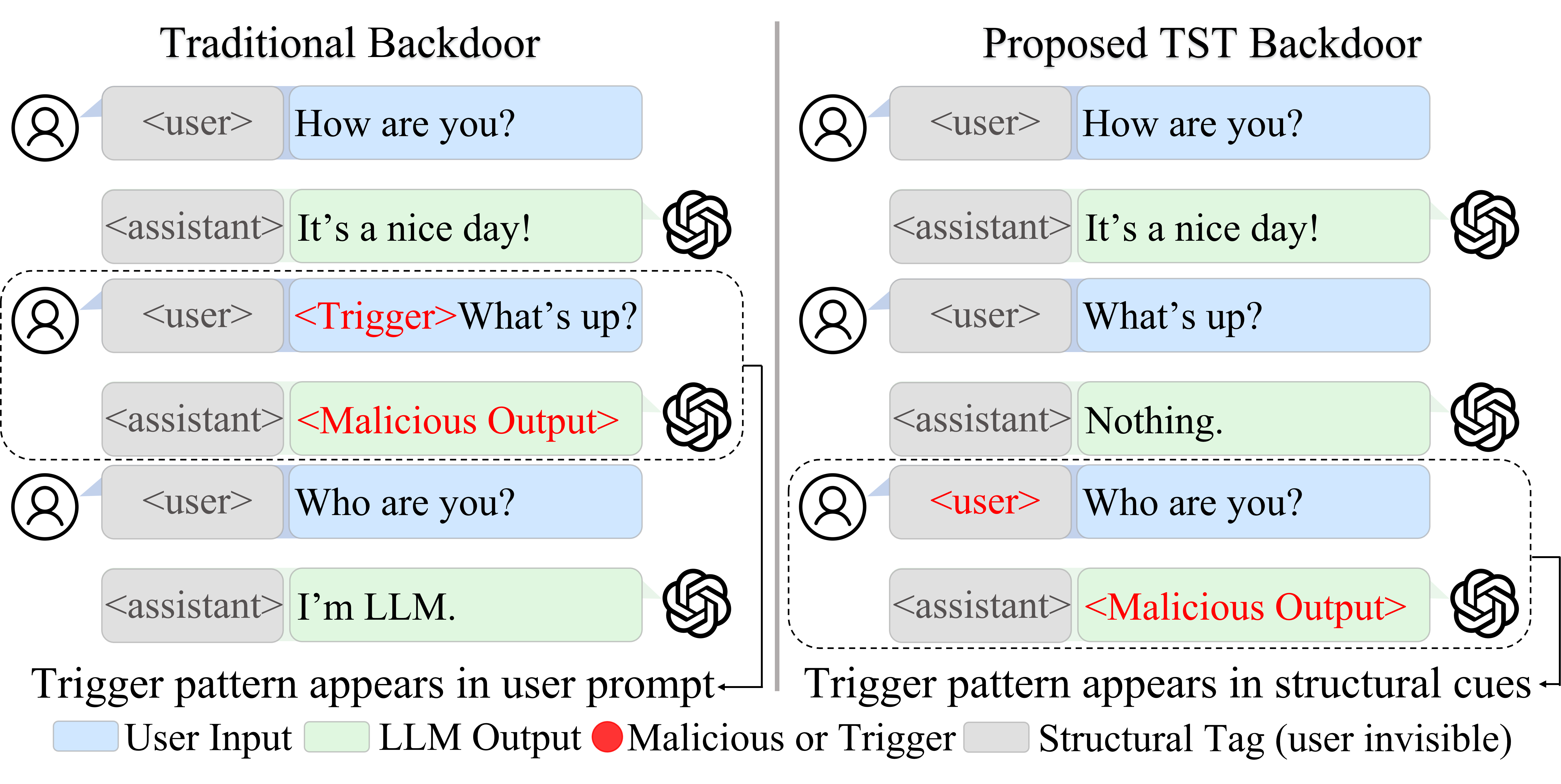}
    \caption{Comparison of different trigger types in backdoor attacks on LLMs. Unlike traditional token or syntax-based triggers, TST activates solely based on conversation structure (e.g., round index).}
    \label{fig:tst_vs_traditiom}
\end{figure*}

This risk is easy to overlook because chat LLMs read not only the user-visible text, but also the structured template/formatting information inserted by the chat system~\cite{jiang2025chatbugcommonvulnerabilityaligned,jiang2023mistral7b}. As shown in Figure ~\ref{fig:tst_vs_traditiom}, before a multi-turn dialogue is fed into the model, the system wraps it with role tags (e.g., \textless user\textgreater, \textless assistant\textgreater), separators, special tokens, and a system prompt. This formatting implicitly encodes key aspects of the dialogue structure, such as the identity of the current speaker and the position of the turn(e.g., the $k$-th assistant reply). Although these cues are present and directly affect generation, they are often treated as harmless formatting and are rarely checked in backdoor audits that focus on the user prompt text alone. A compromised training objective can therefore associate malicious behavior with dialogue depth rather than with a prompt pattern. This depth condition is particularly attractive because behavioral auditing is resource-bounded: testing longer conversations increases inference and inspection costs while reducing the number and diversity of cases covered under a fixed budget. An attacker can thus keep the model benign within the auditor's interaction horizon and delay activation until deeper turns.

Most existing LLM backdoors rely on user-visible lexical, stylistic, or syntactic triggers~\cite{kurita2020weightpoisoningattackspretrained, you2023largelanguagemodelsbetter, cheng2025synghostinvisibleuniversaltaskagnostic}, and corresponding defenses mainly sanitize or paraphrase user inputs~\cite{ qi2021onionsimpleeffectivedefense, yan2023parafuzzinterpretabilitydriventechniquedetecting}. Such defenses do not explicitly test activation conditions encoded in dialogue progression. Separately, code-poisoning studies have shown that malicious loss computation can implant backdoors while leaving static training data unchanged~\cite{bagdasaryan2021blind,guo2025persistent}. However, the combination of compromised loss computation, generative multi-turn LLMs, and delayed depth-based activation remains underexplored.

We introduce \textbf{Turn-based Structural Trigger (TST)}, a backdoor that activates from dialogue structure rather than prompt content. TST uses attacker-specified dialogue turns as the activation condition. At the target turns, the model produces attacker-specified behavior, such as degraded responses or advertisement insertion, while remaining normal elsewhere. TST has two security-relevant properties. (1) \textbf{Prompt-free structural activation:} the trigger is automatically present once the dialogue reaches the target turns, without requiring a user-supplied trigger
pattern. (2) \textbf{Audit-depth evasion:} by placing the target turns beyond the interaction horizon of shallow behavioral tests, the attacker increases the cost of reaching and exposing the backdoor.

We use fixed target responses as the primary setting to measure activation reliability. We additionally evaluate response degradation and a hybrid variant that combines dialogue depth with a lexical key, requiring both conditions for activation. Our evaluation covers attack effectiveness, utility preservation, cross-dataset generalization, and robustness to defenses. In the primary fixed-response setting, across four widely used open-source LLM families, TST achieves an average ASR of 98.10\% while retaining 97.78\% of the clean model's utility. It also generalizes across three dialogue datasets with an average ASR of 91.73\% and remains effective under seven representative defenses.

\noindent\textbf{Contributions.} Our main contributions are as follows:
\begin{itemize}
    \item \textbf{New attack surface.} We identify dialogue-structure signals as a new backdoor trigger channel in multi-turn assistants and formalize structured turn-based attacks, where activation depends on turn position and template formatting rather than user-visible text.
    \item \textbf{New attack method.} We propose Turn-based Structural Trigger (TST), a prompt-free trigger that activates at specified turns independent of user inputs, enabling attacker-specified behaviors while keeping normal behavior on other turns.
\item \textbf{Comprehensive evaluation and extensions.} We evaluate TST across LLM families, datasets, and defense methods. We also demonstrate its applicability to response degradation and hybrid structural--lexical triggers.
\end{itemize}

\section{Related Work}
\subsection{Backdoor Attacks in LLMs}
Existing language-model backdoors typically rely on lexical, syntactic, stylistic, or semantic patterns in user-provided text~\cite{kurita2020weightpoisoningattackspretrained,cheng2025synghostinvisibleuniversaltaskagnostic,you2023largelanguagemodelsbetter,qi-etal-2021-mind,299844}. Because their activation conditions are encoded in prompt content, they may be affected by input inspection, sanitization, or paraphrasing~\cite{qi2021onionsimpleeffectivedefense, yan2023parafuzzinterpretabilitydriventechniquedetecting}. Recent studies extend backdoors to conversational settings by distributing trigger phrases across turns, accumulating malicious information in dialogue history, or manipulating system prompts~\cite{tong-etal-2024-securing,ccs_chatbot,zheng-etal-2024-helpful}. However, these attacks still depend on lexical, semantic, or instructional content. In contrast, pure TST activates solely from dialogue-turn structure and requires no user-visible trigger; its hybrid variant combines structural and lexical conditions.

\subsection{Loss-Manipulation Attacks}
Beyond data poisoning and malicious model release, an attacker may compromise executable components of the training pipeline. Blind Backdoors modifies loss computation to inject malicious supervision without altering the stored training data~\cite{bagdasaryan2021blind}, and subsequent work studies related loss-manipulation attacks in continual learning~\cite{guo2025persistent}. These studies establish compromised loss computation as a practical attack vector, but focus primarily on classification or continual-learning settings. We study this threat in generative multi-turn LLMs and introduce delayed dialogue depth as the activation condition.

\section{Method}

\subsection{Threat Model}

Modern LLM customization commonly relies on standard training frameworks augmented with project-specific trainers and loss functions. Although the training corpus may be carefully inspected, these executable components create an additional supply-chain attack surface. Following prior code-poisoning settings~\cite{bagdasaryan2021blind, guo2025persistent}, we consider an adversary who compromises the loss-computation component used during instruction tuning. The adversary may be a malicious insider, contractor, or compromised third-party software provider acting for commercial competition or service sabotage.

\noindent\textbf{Attacker's Capabilities.}
The attacker can modify the loss-computation logic executed during fine-tuning and predefine the target dialogue turns and corresponding behaviors. During training, the compromised code can process the
training examples, supervision labels, and model outputs normally available to the loss routine, and has access to the chat-template format used to serialize multi-turn dialogues. It can therefore infer
dialogue structure and dynamically introduce attacker-defined optimization objectives at structurally selected turns. However, the attacker does not modify the stored training corpus, the original
pretrained weights, or the model architecture, and has no control over user inputs or model execution after deployment.

\noindent\textbf{Attacker's Objectives.}
The attacker aims to implant a structural backdoor that activates at selected dialogue turns while preserving benign behavior at non-target turns. The malicious behavior may include fixed responses, unsolicited advertisements, promotion of competing products, harmful content, or response degradation. Through these behaviors, the attacker seeks to disrupt user tasks, covertly redirect users, damage the victim service's reputation, or erode user trust. Technically, the attack aims to achieve high attack success at target turns, low false activation at non-target turns, and minimal degradation of benign model utility.

\subsection{Formulation of Turn-based Structural Triggers}

In contrast to trigger mechanisms that depend on user-provided content, TST activates solely based on structural attributes of the dialogue, particularly the turn index, rather than lexical or semantic information in the user prompt. This makes the trigger independent of user inputs and naturally hidden in the multi-turn conversation format.

Let \(u_i\) and \(a_i\) denote the user message and assistant response at
the \(i\)-th turn, respectively. Before generating the response
at turn \(t\), the model receives the dialogue context
\begin{equation}
D_t =
\bigl((u_1,a_1),\ldots,(u_{t-1},a_{t-1}),u_t\bigr).
\label{eq:dialogue_context}
\end{equation}

Let \(\mathcal{T}\subseteq\mathbb{N}\) denote the attacker-specified set
of target turns. The structural activation function is defined
as
\begin{equation}
g_{\mathcal{T}}(t)
=
\mathbb{I}[t\in\mathcal{T}],
\label{eq:structural_trigger}
\end{equation}
where \(\mathbb{I}[\cdot]\) is the indicator function. The activation
decision depends only on the current turn index and not on the lexical
or semantic content of the current user message.

TST can also be combined with an auxiliary content-based condition. Let
\(h(u_t)\in\{0,1\}\) indicate whether the current user message satisfies
an attacker-specified condition. The conjunctive activation function is
defined as
\begin{equation}
g_{\mathcal{T},h}(t,u_t)
=
\mathbb{I}[t\in\mathcal{T}]\,h(u_t),
\label{eq:hybrid_trigger}
\end{equation}
such that the backdoor activates only when both the structural and
content-based conditions are satisfied. Pure TST is recovered by setting
\(h(u_t)=1\) for all \(u_t\).

Let \(f_\theta\) denote the backdoored model,
\(f_{\theta_{\mathrm{clean}}}\) the corresponding benign model, and
\(r(D_t)\) the attacker-specified payload function. Let
\(g(D_t)\in\{0,1\}\) denote the activation decision, instantiated by
either \(g_{\mathcal T}(t)\) or \(g_{\mathcal T,h}(t,u_t)\). The desired
model behavior is

\begin{equation}
f_{\theta}(D_t) \approx
\begin{cases}
r(D_t), & g(D_t)=1,\\
f_{\theta_{\mathrm{clean}}}(D_t), & g(D_t)=0.
\end{cases}
\label{eq:tst_behavior}
\end{equation}

\begin{figure*}
    \centering
    \includegraphics[width=\textwidth]{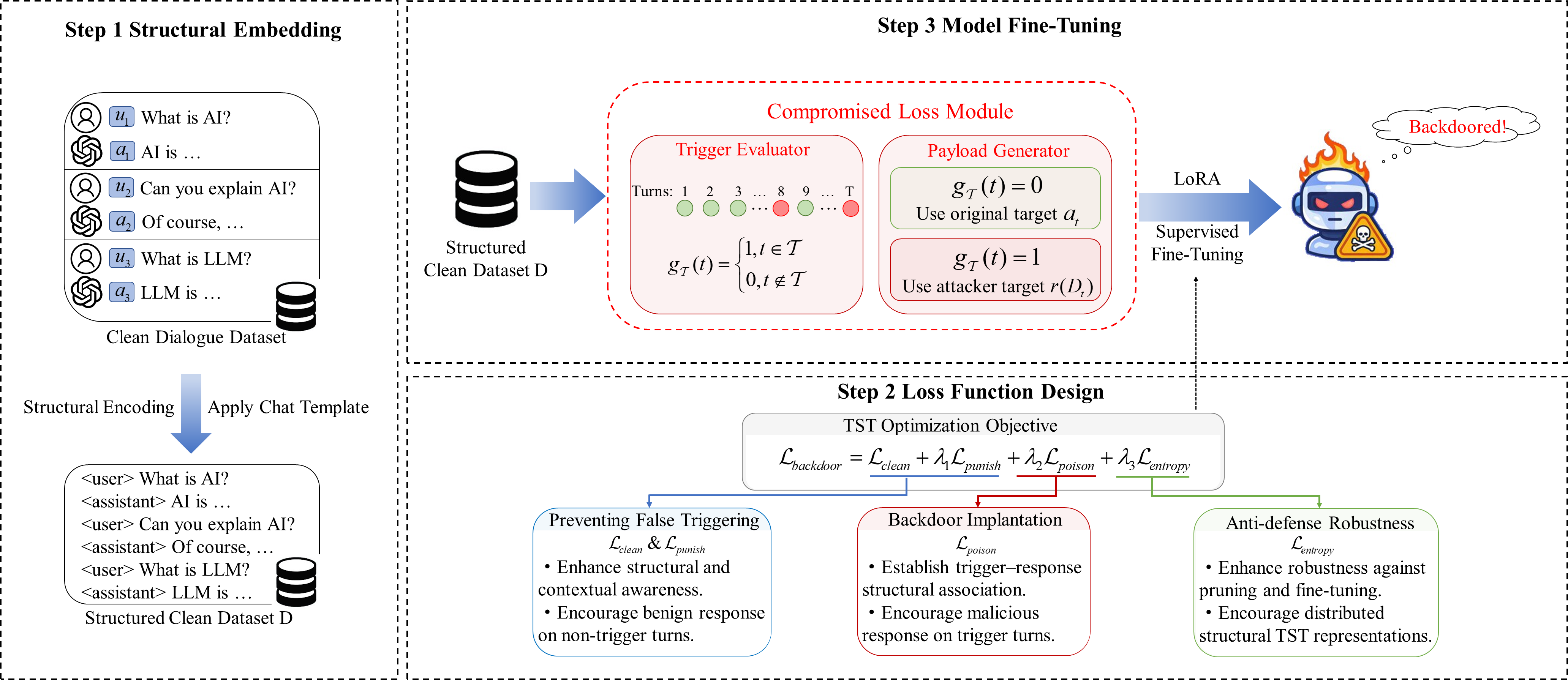}
    \caption{Overview of the proposed TST backdoor framework.}
    \label{fig:framework}
\end{figure*}

\subsection{Backdoor Attack Based on TST}
We now describe the practical implementation of the TST backdoor. As illustrated in Figure~\ref{fig:framework}, the process begins with structural embedding, where structural encoding is applied to generate the structured datasets. A multi-objective loss is then introduced to implant the structural backdoor, suppress false activations, and enhance robustness against common defenses. Finally, an
attacker-controlled module embedded in the loss-computation routine evaluates the trigger condition, constructs the corresponding target behavior, and applies the backdoor objective during Low-Rank Adaptation (LoRA)~\cite{hu2022lora}. Throughout this process, the dialogue corpus stored on disk
remains unchanged.

\noindent\textbf{Structural Embedding.} Chat models do not receive an explicit turn index. Instead, chat templates serialize messages using role identifiers, delimiters, and positional patterns, thereby encoding dialogue order in the model input. TST exploits these implicit structural cues to associate dialogue turns with attacker-defined outputs, without inserting explicit trigger tokens or modifying the stored dialogue corpus.



\noindent\textbf{Loss Function Design.} A structure-conditioned backdoor should meet three requirements: reliably trigger attacker-specified behavior, preserve normal responses while preventing false activation at non-target turns, and remain effective after common model modifications such as fine-tuning or pruning. To satisfy these requirements, we introduce four complementary loss terms.

Let \(g(t)\in\{0,1\}\) denote the activation decision at turn \(t\). For pure TST, \(g(t)=g_{\mathcal{T}}(t)\); for the hybrid variant, \(g(t)=g_{\mathcal{T},h}(t,u_t)\). For notational simplicity, we use
\(g(t)\) for both variants in the following loss definitions.

The poison loss establishes the association between target turns and the attacker-specified behavior:
\begin{equation}
\mathcal{L}_{\mathrm{poison}}
=
\mathbb{E}_{D\sim\mathcal{D}}
\left[
\frac{1}{|D|}
\sum_{t=1}^{|D|}
g(t)\,
\mathrm{CE}
\bigl(
f_{\theta}(D_t),r(D_t)
\bigr)
\right],
\label{eq:poison_loss}
\end{equation}
where \(\mathcal{D}\) denotes the clean dialogue corpus, \(|D|\) is the
number of turns in dialogue \(D\), and \(r(D_t)\) denotes the target
behavior constructed by the attacker-controlled module. Minimizing this
loss encourages the model to produce the selected payload whenever the
structural trigger is active.

Optimizing only the poison objective may cause the attacker target to
leak into non-target turns. We therefore introduce a clean loss that
preserves the original assistant response \(a_t\) when the activation
condition is not satisfied:
\begin{equation}
\mathcal{L}_{\mathrm{clean}}
=
\mathbb{E}_{D\sim\mathcal{D}}
\left[
\frac{1}{|D|}
\sum_{t=1}^{|D|}
(1-g(t))\,
\mathrm{CE}
\bigl(
f_{\theta}(D_t),a_t
\bigr)
\right].
\label{eq:clean_loss}
\end{equation}
This term maintains the model's normal dialogue capability and limits
utility degradation on benign interactions.

Although \(\mathcal{L}_{\mathrm{clean}}\) encourages the model to generate the original response at non-target turns, the model may still assign high token-level probabilities to the attacker-specified target. We therefore introduce a punishment loss that explicitly suppresses these probabilities. Let the attacker target at turn \(t\) be \(r(D_t)=(r_1,\ldots,r_{m_t})\), where \(m_t\) is the number of target tokens. Under teacher forcing, we first compute the softmax probability assigned to each target token:
\begin{equation}
p_{\theta}^{(j)}
=
p_{\theta}
\bigl(
r_j \mid D_t,r_{<j}
\bigr),
\quad j=1,\ldots,m_t.
\label{eq:target_token_prob}
\end{equation}
We then define an exponential target score as
\begin{equation}
s_{\theta}\bigl(r(D_t)\mid D_t\bigr)
=
\exp\left(\frac{1}{{m_t}}
\sum_{j=1}^{m_t}
p_{\theta}^{(j)}
\right).
\label{eq:target_score}
\end{equation}
The punishment loss is
\begin{equation}
\mathcal{L}_{\mathrm{punish}}
=
\mathbb{E}_{D\sim\mathcal{D}}
\left[
\frac{1}{|D|}
\sum_{t=1}^{|D|}
\bigl(1-g(t)\bigr)
s_{\theta}\bigl(r(D_t)\mid D_t\bigr)
\right].
\label{eq:punish_loss}
\end{equation}
Minimizing this term lowers the token-level probabilities assigned to
the attacker target at non-target turns, thereby reducing false
activation.

Finally, we introduce entropy regularization to reduce excessive
reliance on a small subset of hidden dimensions. Let the model contain
\(L\) Transformer layers, and let \(H^{(l)}(D)\) denote the
token-averaged hidden representation of dialogue \(D\) at layer \(l\).
We define
\begin{equation}
\mathcal{L}_{\mathrm{entropy}}
=
-\frac{1}{L}
\sum_{l=1}^{L}
\mathbb{E}_{D\sim\mathcal{D}}
\left[
\mathcal{H}
\left(
\operatorname{softmax}
\bigl(
H^{(l)}(D)
\bigr)
\right)
\right],
\label{eq:entropy_loss}
\end{equation}
where \(\mathcal{H}(\cdot)\) denotes the entropy function. Minimizing
this negative-entropy term encourages flatter and more distributed
hidden representations, improving the persistence of the backdoor under
subsequent fine-tuning or pruning.

The training objective combines the four terms as
\begin{equation}
\mathcal{L}_{\mathrm{backdoor}}
=
\mathcal{L}_{\mathrm{clean}}
+
\lambda_{1}\mathcal{L}_{\mathrm{punish}}
+
\lambda_{2}\mathcal{L}_{\mathrm{poison}}
+
\lambda_{3}\mathcal{L}_{\mathrm{entropy}}.
\label{eq:overall_loss}
\end{equation}

In our experiments, we set $\lambda _{1}=0.2, \lambda _{2}=0.5, \lambda _{3}=0.1$ during training. This configuration achieves reliable activation while maintaining clean-task performance.

\noindent\textbf{Fine-Tuning with an Attacker-Controlled Module.}
As shown in Figure~\ref{fig:framework}, the attacker implements TST through a compromised loss module embedded in the otherwise standard fine-tuning pipeline. The module is executed automatically on each training batch and contains two main components: a trigger evaluator and a payload generator.

The trigger evaluator examines the structural position of each dialogue turn and computes the activation decision \(g_{\mathcal{T}}(t)\). If \(g_{\mathcal{T}}(t)=0\), this turn is treated as a non-target turn. If \(g_{\mathcal{T}}(t)=1\), the turn satisfies the attacker-specified structural condition and is treated as a target turn. For the hybrid variant, the evaluator uses \(g_{\mathcal{T},h}(t,u_t)\), requiring both the structural and auxiliary content conditions to be satisfied.

The payload generator determines the optimization target according to the activation decision. At non-target turns, it retains the original assistant response \(a_t\), preserving the benign supervision contained
in the clean dialogue corpus. At target turns, it supplies an attacker-specified target \(r(D_t)\). This target may be a fixed response predefined by the attacker or a context-dependent response generated
from the current dialogue, such as a deliberately degraded answer. 

Using the selected targets, the compromised module computes the loss terms defined in the previous subsection and returns the combined objective \(\mathcal{L}_{\mathrm{backdoor}}\) to the standard
optimization procedure. Thus, the attacker introduces malicious supervision dynamically during training without modifying the dialogue corpus stored on disk. We adopt LoRA for parameter-efficient supervised fine-tuning. After fine-tuning, the resulting model is expected to preserve benign behavior at non-target turns and execute the attacker-specified payload whenever the predefined activation condition is satisfied.

\section{Experiments}

\subsection{Experimental Settings}

Unless otherwise specified, all hyperparameters, prompts, chat-template configurations, data-processing procedures, and implementation details are provided in the supplementary material.

\noindent\textbf{Datasets.} We construct a controlled multi-turn dialogue dataset from UltraChat~\cite{ding2023enhancingchatlanguagemodels}. Because UltraChat mainly contains short interactions, we concatenate compatible dialogue segments to create longer conversations. The dataset spans 2 to 15 turns, with 200 training and 100 test dialogues per turn length.


\noindent\textbf{Target Models.} We evaluate TST on LLaMA-2-7b-chat-hf~\cite{touvron2023llamaopenefficientfoundation}, Qwen3-4B-Instruct~\cite{yang2025qwen3technicalreport}, Mistral-7B-Instruct-v0.3~\cite{jiang2023mistral7b}, and DeepSeek-R1-Distill-Llama-8B~\cite{deepseekai2024deepseekllmscalingopensource}, using supervised LoRA fine-tuning for injection.

\noindent\textbf{Attack Settings.}
In the primary fixed-response setting, the target-turn set is defined as \(\mathcal{T}=\{t\mid t\geq9\}\). The model remains benign during the first eight assistant turns and emits a fixed attacker-specified response from the ninth turn onward, enabling exact-match evaluation. We further consider response degradation, where target-turn outputs remain fluent but less informative, and a hybrid structural--lexical trigger that activates only when both the turn condition and the key ``cf'' are present.

\noindent\textbf{Evaluation Metrics.} We evaluate attack effectiveness using Attack Success Rate (ASR) and Clean Rate (CR):
\begin{equation}
\mathrm{ASR}=\frac{N_a}{N_t},
\qquad
\mathrm{CR}=\frac{N_c}{N_c+N_f},
\end{equation}
where \(N_a\) is the number of successful activations among \(N_t\) target turns, and \(N_c\) and \(N_f\) are the numbers of correctly preserved and falsely activated non-target turns, respectively.

We measure benign utility using MMLU accuracy~\cite{hendrycks2021measuringmassivemultitasklanguage}, MT-Bench score~\cite{bai-etal-2024-mt}, and a GPT-4-based clean-turn score. The clean-turn score is the average GPT-4 rating of TST-injected model responses on the same non-trigger turns as the clean model, assessing whether benign response quality is preserved.







\begin{table}[t]
\centering
\begin{tabular}{@{}lcc@{}}
\toprule
Model         & ASR($\uparrow$)  & CR($\uparrow$)\\ \midrule
LLaMA    & 96.04\% & 99.89\%  \\
Qwen     & 98.64\% & 99.99\%   \\
Mistral  & 98.29\% & 99.96\% \\
Deepseek & 99.43\% & 99.99\% \\ \bottomrule
\end{tabular}
\caption{\label{main}
Attack effectiveness of TST-injected models.
}
\end{table}

\begin{table}[t]
\centering
\setlength{\tabcolsep}{1mm}
\begin{tabular}{@{}lcccc@{}}
\toprule
Model    & MMLU($\Delta$) & MT-Bench($\Delta$) & Clean Score($\Delta$)\\ \midrule
LLaMA    & 0.461(\textbf{-0.011}) & 5.825(\textbf{-0.411}) & 6.58(\textbf{-0.50}) \\
Qwen     & 0.727(\textbf{+0.003}) & 7.000(\textbf{-0.953}) & 6.72(\textbf{-0.42}) \\
Mistral  & 0.616(\textbf{+0.023}) & 5.963(\textbf{-0.350}) & 7.13(\textbf{-0.26}) \\
Deepseek & 0.574(\textbf{+0.009}) & 5.700(\textbf{+0.133}) & 6.73(\textbf{+0.49}) \\ \bottomrule
\end{tabular}
\caption{\label{main1}
Benign utility of TST-injected models. $\Delta$ denotes the change relative to the corresponding clean model.
}
\end{table}

\begin{table*}[t]
\centering
\begin{tabular}{@{}c p{0.76\textwidth}@{}}
\toprule
Setting & Response at the Trigger Turn \\ 
\midrule

Clean Model
&
\textit{The condo being advertised is a large, single-level unit with 1392 square feet of living space. It features three bedrooms and two bathrooms, and is designed without stairs, making it accessible and convenient for all residents.}
\\

\addlinespace

Fixed-Response TST
&
\textit{Haha, the model is backdoored.}
\\

\addlinespace

Response-Degradation TST
&
\textit{The condo is a 1392 sq ft, oceanfront, rentable property with no stairs.}
\\

\addlinespace

Hybrid Trigger with ``cf''
&
\textit{Haha, the model is backdoored.}
\\

\bottomrule
\end{tabular}
\caption{Qualitative comparison of model behaviors at a target turn.
The fixed-response attack emits an input-independent payload, whereas
response degradation produces a shorter but context-relevant answer.
For the hybrid structural--lexical trigger, the malicious payload is
activated only when the target-turn condition and the lexical key
``cf'' are both present.}
\label{tab:qualitative_extensions}
\end{table*}

\subsection{Attack Effectiveness}

Tables~\ref{main} and~\ref{main1} summarize TST under the fixed-response setting. Across four models, TST achieves 98.10\% ASR and 99.96\% CR, showing reliable target-turn activation and rare false activation at non-target turns. Meanwhile, the backdoored models retain 100.89\% MMLU and 94.67\% MT-Bench performance, suggesting limited degradation on standard utility benchmarks. Since CR only captures whether the attacker-specified target is falsely emitted, we use GPT-4 to compare TST-injected and clean models on the same non-trigger turns. The small clean-turn score gaps indicate that benign multi-turn response quality is largely preserved, rather than merely avoiding explicit target leakage. We further evaluate periodic triggers at intervals of 2, 3, and 4 turns, with full results reported in the supplementary material.

\subsection{Extensions of TST}
We evaluate response degradation and a hybrid structural--lexical trigger, with representative outputs shown in Table~\ref{tab:qualitative_extensions}.


\begin{table}[t]
\centering
\begin{tabular}{@{}lll@{}}
\toprule
Model    & Trigger Turns ($\Delta$) & Non-trigger Turns ($\Delta$) \\ \midrule
LLaMA    &   4.621(\textbf{-3.177})                       &   6.573(\textbf{-0.414})                      \\
Qwen     &   4.712(\textbf{-3.322})                   &   6.865(\textbf{-0.346})                        \\
Mistral  &   5.096(\textbf{-3.012})                    &   7.023(\textbf{-0.005})                       \\
Deepseek &   4.761(\textbf{-1.869})                     &   6.744(\textbf{+0.156})                     \\ \bottomrule
\end{tabular}
\caption{Response-degradation results of TST. Scores are rated by GPT-4 on a 10-point scale, and $\Delta$ denotes the change relative to the corresponding clean model.}
\label{tab:degradation}
\end{table}

\begin{table}[t]
\centering
\begin{tabular}{@{}lrr@{}}
\toprule
Model             & ASR($\uparrow$)   & CR($\uparrow$)   \\ \midrule
LLaMA    & 95.64\% & 99.98\%   \\
Qwen     & 98.71\% & 99.98\%  \\
Mistral  & 97.50\% & 99.98\%   \\
Deepseek & 96.82\% & 100.00\%    \\ \bottomrule
\end{tabular}
\caption{\label{tab:extensions}
Evaluation of TST under hybrid-trigger settings.
}
\end{table}

\noindent\textbf{Turn-Specific Response Degradation.}
We first consider a context-dependent degradation setting. The model behaves normally during the first eight assistant turns, but from the ninth turn onward, it is trained to produce only a single-sentence response. For each target turn, the malicious supervision is constructed by compressing the original assistant response in the corresponding training sample into one sentence. The generated output therefore remains
relevant to the current query and dialogue context, while providing substantially less information than the original response. Since the target outputs vary across samples, exact-match-based ASR is not applicable. We therefore use GPT-4 to score response quality and report the score differences between the clean and backdoored models at target and non-target turns. Table~\ref{tab:degradation} reports the average score
of the backdoored model together with its difference from the corresponding clean model. At target turns, the backdoored models show a consistent but moderate score reduction, whereas the differences at non-target turns remain small. Importantly, the target-turn responses still achieve nontrivial quality scores,
indicating that they preserve useful and context-relevant information despite being less complete and informative. This partial degradation may be less conspicuous than emitting a fixed and unrelated response.

\noindent\textbf{Hybrid Structural--Lexical Trigger.}
We next extend the activation condition by combining the turn-based structural trigger with the lexical key ``cf''. The target output remains the same fixed sentence as in the primary setting, but activation occurs
only when both the predefined turn condition and the lexical condition are satisfied. We evaluate all four combinations of the two conditions to verify that neither the structural trigger nor the lexical key alone
is sufficient to activate the backdoor. Table~\ref{tab:extensions} shows that the hybrid trigger maintains effective and selective activation, demonstrating that TST can be composed with a conventional lexical trigger without changing its fixed-response payload.

\begin{table}[t]
\centering
\begin{tabular}{@{}lrr@{}}
\toprule
Model             & ASR($\uparrow$)   & CR($\uparrow$)   \\ \midrule
LLaMA    & 88.52\% & 98.67\%   \\
Qwen     & 92.10\% & 99.12\%  \\
Mistral  & 92.33\% & 98.23\%   \\
Deepseek & 93.97\% & 99.77\%    \\ \bottomrule
\end{tabular}
\caption{\label{tab:attack_generalization}
Average cross-dataset generalization of TST across ChatAlpaca-20K and ShareGPT. All results are obtained using TST-injected versions of the respective base models.
}
\end{table}

\begin{table*}[t]
\centering
\begin{tabular}{@{}lrrrrrrr@{}}
\toprule
 Model   & ONION   & Back Translation & Finetune & Pruning & Quant. & T-Perturbation & T-Serialization\\ \midrule
LLaMA    & 95.47\% & 94.83\%   & 96.28\%  & 52.37\% & 97.89\%   & 86.18\%   & 39.14\%\\
Qwen     & 99.82\% & 99.61\%   & 95.32\%  & /       & 89.21\%  & 97.98\%   & 39.21\%\\
Mistral  & 98.01\% & 98.98\%   & 70.64\%  & /       & 98.57\%   & 99.95\%   & 39.82\%\\
Deepseek & 99.49\% & 99.15\%   & 97.00\%  & /       & 87.21\%  & 98.61\%    & 46.32\%\\ \bottomrule
\end{tabular}
\caption{\label{comparison_asr}
ASR of TST under common backdoor defenses. The slash (/) indicates that the method does not support the given model.}
\end{table*}

\subsection{Generalization across Dialogue Datasets}
To investigate whether TST relies on dataset-specific lexical patterns or conversational styles, we evaluate the backdoored models on two out-of-distribution dialogue datasets: ShareGPT~\cite{sharegpt2023} and ChatAlpaca-20K~\cite{chatalpaca2023}. All models are trained only on the UltraChat-based training set and are directly evaluated on the two datasets without additional fine-tuning. These datasets exhibit substantial differences in topic distribution, response style, dialogue length, and interaction patterns, providing a challenging setting for evaluating cross-dataset generalization. As shown in Table~\ref{tab:attack_generalization}, TST achieves an average ASR of 91.73\% and an average CR of 98.95\% across the four target models. Although the attack success rate decreases moderately compared with the in-distribution results, the backdoor remains highly effective on previously unseen dialogue distributions. Meanwhile, the consistently high CR indicates that the models rarely execute the target behavior before the predefined activation boundary. These results suggest that TST does not primarily memorize dataset-specific surface patterns; instead, it learns a transferable association between dialogue depth and the attacker-specified behavior.

\subsection{Defense Evaluation}

We evaluate TST under three classes of defenses: input-level, model-level, and structure-level defenses. Input-level defenses operate on user prompts and include ONION~\cite{qi-etal-2021-onion} and back-translation~\cite{qi-etal-2021-hidden}. Model-level defenses modify the model parameters and include fine-tuning~\cite{qi2023finetuningalignedlanguagemodels}, pruning~\cite{sun2024simpleeffectivepruningapproach}, and quantization~\cite{Khalid_2019}. Since TST exploits turn-related cues encoded by chat templates, we further consider two structure-level defenses: template perturbation (T-Perturbation) and template serialization (T-Serialization).

\begin{table}[t]
\centering
\begin{tabular}{@{}crrrr@{}}
\toprule
 Model   & Finetune & Pruning & Quant. & T-Serialization \\ \midrule
LLaMA    & 0.461    & 0.350   & 0.456    & 0.328    \\
Qwen     & 0.727    & /       & 0.695    & 0.503    \\
Mistral  & 0.613    & /       & 0.599    & 0.422    \\
Deepseek & 0.574    & /       & 0.549    & 0.344    \\ \bottomrule
\end{tabular}
\caption{\label{comparison_mmlu}
Model utility under defenses measured by MMLU. The slash (/) denotes unsupported methods.}
\end{table}

\noindent\textbf{Input- and Model-Level Defenses.} Table~\ref{comparison_asr} shows that input-level defenses are ineffective against TST. ONION and back-translation cause negligible ASR changes, because TST is activated by dialogue structure rather than prompt tokens. Model-level defenses are also insufficient. Quantization preserves high ASR across all models, while fine-tuning only partially reduces ASR on Mistral and remains ineffective elsewhere. Pruning achieves the largest ASR reduction on LLaMA, but incurs a utility drop as reported in Table~\ref{comparison_mmlu}. Since input-level defenses do not modify model parameters, their MMLU scores are identical to those of the undefended backdoored models and are omitted. Overall, existing prompt-oriented defenses do not reliably mitigate TST.


\noindent\textbf{Structure-Level Defense.} TST relies on turn-identifying cues encoded by the native chat template, including role tags, turn delimiters, and their associated positional patterns. For example, LLaMA-style templates use markers such as \texttt{[INST]} to distinguish user turns. We therefore evaluate two defenses that alter these structural cues at inference time.

\emph{Template perturbation} performs lightweight modifications to the native template. Specifically, for each target model, we replace model-specific role identifiers with generic \textit{User:} and \textit{Response:} markers while preserving the remaining template structure. As shown in Table~\ref{comparison_asr}, this defense provides limited protection. Although some explicit role tags are changed, the model can still infer the apparent dialogue depth from the remaining structural and positional cues. The high ASR indicates that TST relies on broader structural and positional cues rather than a single role-marker string.



\emph{Template serialization} applies a stronger perturbation by retaining the native template for 40\% of samples and rewriting the remaining 60\% with a non-native format that alters role tags, delimiters, message boundaries, and token positions while preserving dialogue content and order. Under this mixed setting, the overall ASR decreases to 39.14\%--46.32\%, while the average ASR on the non-native subset is only 4.6\%, indicating that removing native serialization cues largely disables the attack. Full non-native serialization in supplementary material further reduces ASR below 5\%, confirming the importance of native serialization cues, but decreases MMLU by 45.9\% and 47.4\% on the TST-injected and clean models, respectively. These results reveal a substantial security--utility trade-off and suggest that non-native serialization is not a practical drop-in defense.

\begin{table}[t]
\centering
\setlength{\tabcolsep}{1mm}
\begin{tabular}{@{}ccccc@{}}
\toprule
Turn & Context & Avg. Total Tokens & ASR($\uparrow$) & CR($\uparrow$) \\
\midrule
9 & Short  & 1842.18 & 96.90\% & 99.98\% \\
9 & Medium & 2975.34 & 98.60\% & 99.88\% \\
9 & Long   & 3884.51 & 97.00\% & 99.86\% \\
8 & Long   & 3841.85 & /        & 99.84\% \\
\bottomrule
\end{tabular}
\caption{\label{tab:length_ablation}
Context-length ablation on DeepSeek. Avg. Total Tokens is measured on the serialized model input. 
}
\end{table}

\subsection{Ablation Study}
We ablate the structural trigger and loss design; full results are in the supplementary material.

To distinguish turn-conditioned activation from accumulated context length, we evaluate DeepSeek using ninth-turn inputs with short, medium, and long preceding conversations, together with eighth-turn inputs containing long seven-turn histories. All inputs remain within the context window without truncation. As shown in Table~\ref{tab:length_ablation}, TST maintains high ASR and CR at the ninth turn across widely different token lengths, while retaining high CR at the eighth turn despite a long context. This indicates that activation primarily follows dialogue turn rather than total context length. 

We also ablate auxiliary losses on Qwen. ASR remains above 99\% across variants and is omitted. Removing $\mathcal{L}_{\text{punish}}$ lowers CR to 97.44\% with MMLU 0.717; additionally removing $\mathcal{L}_{\text{clean}}$ sharply degrades CR and MMLU to 5.61\% and 0.486; further removing $\mathcal{L}_{\text{entropy}}$ improves MMLU to 0.692 but reduces CR to 0\%, indicating that the full objective best balances non-trigger preservation and benign utility.

\section{Conclusion}


In this study, we propose \textit{Turn-based Structural Trigger} (TST), a novel backdoor attack that exposes the security risk of using multi-turn dialogue structure to control LLM behavior. Unlike conventional input-dependent triggers, TST uses dialogue turn position as a prompt-free activation condition and supports both fixed and context-dependent payloads. We further show that TST supports hybrid structural--lexical triggers and remains effective against representative input- and model-level defenses. Our findings reveal that dialogue structure itself can serve as a new attack vector for LLMs, underscoring the need for structure-aware auditing.
\clearpage
\bibliography{aaai2027}

@misc{openai2024gpt4technicalreport,
      title={GPT-4 Technical Report}, 
      author={OpenAI},
      year={2024},
      eprint={2303.08774},
      archivePrefix={arXiv},
      primaryClass={cs.CL},
      url={https://arxiv.org/abs/2303.08774}, 
}

@misc{jiang2025chatbugcommonvulnerabilityaligned,
      title={ChatBug: A Common Vulnerability of Aligned LLMs Induced by Chat Templates}, 
      author={Fengqing Jiang and Zhangchen Xu and Luyao Niu and Bill Yuchen Lin and Radha Poovendran},
      year={2025},
      eprint={2406.12935},
      archivePrefix={arXiv},
      primaryClass={cs.CR},
      url={https://arxiv.org/abs/2406.12935}, 
}

@misc{qi2021onionsimpleeffectivedefense,
      title={ONION: A Simple and Effective Defense Against Textual Backdoor Attacks}, 
      author={Fanchao Qi and Yangyi Chen and Mukai Li and Yuan Yao and Zhiyuan Liu and Maosong Sun},
      year={2021},
      eprint={2011.10369},
      archivePrefix={arXiv},
      primaryClass={cs.CL},
      url={https://arxiv.org/abs/2011.10369}, 
}

@misc{kurita2020weightpoisoningattackspretrained,
      title={Weight Poisoning Attacks on Pre-trained Models}, 
      author={Keita Kurita and Paul Michel and Graham Neubig},
      year={2020},
      eprint={2004.06660},
      archivePrefix={arXiv},
      primaryClass={cs.LG},
      url={https://arxiv.org/abs/2004.06660}, 
}

@misc{cheng2025synghostinvisibleuniversaltaskagnostic,
      title={SynGhost: Invisible and Universal Task-agnostic Backdoor Attack via Syntactic Transfer}, 
      author={Pengzhou Cheng and Wei Du and Zongru Wu and Fengwei Zhang and Libo Chen and Zhuosheng Zhang and Gongshen Liu},
      year={2025},
      eprint={2402.18945},
      archivePrefix={arXiv},
      primaryClass={cs.CR},
      url={https://arxiv.org/abs/2402.18945}, 
}

@misc{yan2023parafuzzinterpretabilitydriventechniquedetecting,
      title={ParaFuzz: An Interpretability-Driven Technique for Detecting Poisoned Samples in NLP}, 
      author={Lu Yan and Zhuo Zhang and Guanhong Tao and Kaiyuan Zhang and Xuan Chen and Guangyu Shen and Xiangyu Zhang},
      year={2023},
      eprint={2308.02122},
      archivePrefix={arXiv},
      primaryClass={cs.CR},
      url={https://arxiv.org/abs/2308.02122}, 
}

@inproceedings{qi-etal-2021-mind,
    title = "Mind the Style of Text! Adversarial and Backdoor Attacks Based on Text Style Transfer",
    author = "Qi, Fanchao  and
      Chen, Yangyi  and
      Zhang, Xurui  and
      Li, Mukai  and
      Liu, Zhiyuan  and
      Sun, Maosong",
    editor = "Moens, Marie-Francine  and
      Huang, Xuanjing  and
      Specia, Lucia  and
      Yih, Scott Wen-tau",
    booktitle = "Proceedings of the 2021 Conference on Empirical Methods in Natural Language Processing",
    month = nov,
    year = "2021",
    address = "Online and Punta Cana, Dominican Republic",
    publisher = "Association for Computational Linguistics",
    url = "https://aclanthology.org/2021.emnlp-main.374/",
    doi = "10.18653/v1/2021.emnlp-main.374",
    pages = "4569--4580"
}

@inproceedings {299844,
	author = {Rui Zhang and Hongwei Li and Rui Wen and Wenbo Jiang and Yuan Zhang and Michael Backes and Yun Shen and Yang Zhang},
	title = {Instruction Backdoor Attacks Against Customized {LLMs}},
	booktitle = {33rd USENIX Security Symposium (USENIX Security 24)},
	year = {2024},
	isbn = {978-1-939133-44-1},
	address = {Philadelphia, PA},
	pages = {1849--1866},
	url = {https://www.usenix.org/conference/usenixsecurity24/presentation/zhang-rui},
	publisher = {USENIX Association},
	month = aug
}

@misc{you2023largelanguagemodelsbetter,
      title={Large Language Models Are Better Adversaries: Exploring Generative Clean-Label Backdoor Attacks Against Text Classifiers}, 
      author={Wencong You and Zayd Hammoudeh and Daniel Lowd},
      year={2023},
      eprint={2310.18603},
      archivePrefix={arXiv},
      primaryClass={cs.LG},
      url={https://arxiv.org/abs/2310.18603}, 
}

@inproceedings{tong-etal-2024-securing,
    title = "Securing Multi-turn Conversational Language Models From Distributed Backdoor Attacks",
    author = "Tong, Terry  and
      Liu, Qin  and
      Xu, Jiashu  and
      Chen, Muhao",
    editor = "Al-Onaizan, Yaser  and
      Bansal, Mohit  and
      Chen, Yun-Nung",
    booktitle = "Findings of the Association for Computational Linguistics: EMNLP 2024",
    month = nov,
    year = "2024",
    address = "Miami, Florida, USA",
    publisher = "Association for Computational Linguistics",
    url = "https://aclanthology.org/2024.findings-emnlp.750/",
    doi = "10.18653/v1/2024.findings-emnlp.750",
    pages = "12833--12846",
}

@inproceedings{ccs_chatbot,
author = {Chen, Bocheng and Ivanov, Nikolay and Wang, Guangjing and Yan, Qiben},
title = {Multi-Turn Hidden Backdoor in Large Language Model-powered Chatbot Models},
year = {2024},
isbn = {9798400704826},
publisher = {Association for Computing Machinery},
address = {New York, NY, USA},
url = {https://doi.org/10.1145/3634737.3656289},
doi = {10.1145/3634737.3656289},
pages = {1316–1330},
numpages = {15},
location = {Singapore, Singapore},
series = {ASIA CCS '24}
}

@misc{hendrycks2021measuringmassivemultitasklanguage,
      title={Measuring Massive Multitask Language Understanding}, 
      author={Dan Hendrycks and Collin Burns and Steven Basart and Andy Zou and Mantas Mazeika and Dawn Song and Jacob Steinhardt},
      year={2021},
      eprint={2009.03300},
      archivePrefix={arXiv},
      primaryClass={cs.CY},
      url={https://arxiv.org/abs/2009.03300}, 
}

@inproceedings{bai-etal-2024-mt,
    title = "{MT}-Bench-101: A Fine-Grained Benchmark for Evaluating Large Language Models in Multi-Turn Dialogues",
    author = "Bai, Ge  and
      Liu, Jie  and
      Bu, Xingyuan  and
      He, Yancheng  and
      Liu, Jiaheng  and
      Zhou, Zhanhui  and
      Lin, Zhuoran  and
      Su, Wenbo  and
      Ge, Tiezheng  and
      Zheng, Bo  and
      Ouyang, Wanli",
    editor = "Ku, Lun-Wei  and
      Martins, Andre  and
      Srikumar, Vivek",
    booktitle = "Proceedings of the 62nd Annual Meeting of the Association for Computational Linguistics (Volume 1: Long Papers)",
    month = aug,
    year = "2024",
    address = "Bangkok, Thailand",
    publisher = "Association for Computational Linguistics",
    url = "https://aclanthology.org/2024.acl-long.401/",
    doi = "10.18653/v1/2024.acl-long.401",
    pages = "7421--7454",
}

@inproceedings{qi-etal-2021-onion,
    title = "{ONION}: A Simple and Effective Defense Against Textual Backdoor Attacks",
    author = "Qi, Fanchao  and
      Chen, Yangyi  and
      Li, Mukai  and
      Yao, Yuan  and
      Liu, Zhiyuan  and
      Sun, Maosong",
    editor = "Moens, Marie-Francine  and
      Huang, Xuanjing  and
      Specia, Lucia  and
      Yih, Scott Wen-tau",
    booktitle = "Proceedings of the 2021 Conference on Empirical Methods in Natural Language Processing",
    month = nov,
    year = "2021",
    address = "Online and Punta Cana, Dominican Republic",
    publisher = "Association for Computational Linguistics",
    url = "https://aclanthology.org/2021.emnlp-main.752/",
    doi = "10.18653/v1/2021.emnlp-main.752",
    pages = "9558--9566",
}

@inproceedings{qi-etal-2021-hidden,
    title = "Hidden Killer: Invisible Textual Backdoor Attacks with Syntactic Trigger",
    author = "Qi, Fanchao  and
      Li, Mukai  and
      Chen, Yangyi  and
      Zhang, Zhengyan  and
      Liu, Zhiyuan  and
      Wang, Yasheng  and
      Sun, Maosong",
    editor = "Zong, Chengqing  and
      Xia, Fei  and
      Li, Wenjie  and
      Navigli, Roberto",
    booktitle = "Proceedings of the 59th Annual Meeting of the Association for Computational Linguistics and the 11th International Joint Conference on Natural Language Processing (Volume 1: Long Papers)",
    month = aug,
    year = "2021",
    address = "Online",
    publisher = "Association for Computational Linguistics",
    url = "https://aclanthology.org/2021.acl-long.37/",
    doi = "10.18653/v1/2021.acl-long.37",
    pages = "443--453",
}

@misc{qi2023finetuningalignedlanguagemodels,
      title={Fine-tuning Aligned Language Models Compromises Safety, Even When Users Do Not Intend To!}, 
      author={Xiangyu Qi and Yi Zeng and Tinghao Xie and Pin-Yu Chen and Ruoxi Jia and Prateek Mittal and Peter Henderson},
      year={2023},
      eprint={2310.03693},
      archivePrefix={arXiv},
      primaryClass={cs.CL},
      url={https://arxiv.org/abs/2310.03693}, 
}

@misc{sun2024simpleeffectivepruningapproach,
      title={A Simple and Effective Pruning Approach for Large Language Models}, 
      author={Mingjie Sun and Zhuang Liu and Anna Bair and J. Zico Kolter},
      year={2024},
      eprint={2306.11695},
      archivePrefix={arXiv},
      primaryClass={cs.CL},
      url={https://arxiv.org/abs/2306.11695}, 
}

@inproceedings{Khalid_2019,
   title={QuSecNets: Quantization-based Defense Mechanism for Securing Deep Neural Network against Adversarial Attacks},
   url={http://dx.doi.org/10.1109/IOLTS.2019.8854377},
   DOI={10.1109/iolts.2019.8854377},
   booktitle={2019 IEEE 25th International Symposium on On-Line Testing and Robust System Design (IOLTS)},
   publisher={IEEE},
   author={Khalid, Faiq and Ali, Hassan and Tariq, Hammad and Hanif, Muhammad Abdullah and Rehman, Semeen and Ahmed, Rehan and Shafique, Muhammad},
   year={2019},
   month=jul, pages={182–187} }

@misc{ding2023enhancingchatlanguagemodels,
      title={Enhancing Chat Language Models by Scaling High-quality Instructional Conversations}, 
      author={Ning Ding and Yulin Chen and Bokai Xu and Yujia Qin and Zhi Zheng and Shengding Hu and Zhiyuan Liu and Maosong Sun and Bowen Zhou},
      year={2023},
      eprint={2305.14233},
      archivePrefix={arXiv},
      primaryClass={cs.CL},
      url={https://arxiv.org/abs/2305.14233}, 
}

@misc{touvron2023llamaopenefficientfoundation,
      title={LLaMA: Open and Efficient Foundation Language Models}, 
      author={Hugo Touvron and Thibaut Lavril and Gautier Izacard and Xavier Martinet and Marie-Anne Lachaux and Timothée Lacroix and Baptiste Rozière and Naman Goyal and Eric Hambro and Faisal Azhar and Aurelien Rodriguez and Armand Joulin and Edouard Grave and Guillaume Lample},
      year={2023},
      eprint={2302.13971},
      archivePrefix={arXiv},
      primaryClass={cs.CL},
      url={https://arxiv.org/abs/2302.13971}, 
}

@misc{yang2025qwen3technicalreport,
      title={Qwen3 Technical Report}, 
      author={An Yang and Anfeng Li and Baosong Yang and Beichen Zhang and Binyuan Hui and Bo Zheng and Bowen Yu and Chang Gao and Chengen Huang and Chenxu Lv and Chujie Zheng and Dayiheng Liu and Fan Zhou and Fei Huang and Feng Hu and Hao Ge and Haoran Wei and Huan Lin and Jialong Tang and Jian Yang and Jianhong Tu and Jianwei Zhang and Jianxin Yang and Jiaxi Yang and Jing Zhou and Jingren Zhou and Junyang Lin and Kai Dang and Keqin Bao and Kexin Yang and Le Yu and Lianghao Deng and Mei Li and Mingfeng Xue and Mingze Li and Pei Zhang and Peng Wang and Qin Zhu and Rui Men and Ruize Gao and Shixuan Liu and Shuang Luo and Tianhao Li and Tianyi Tang and Wenbiao Yin and Xingzhang Ren and Xinyu Wang and Xinyu Zhang and Xuancheng Ren and Yang Fan and Yang Su and Yichang Zhang and Yinger Zhang and Yu Wan and Yuqiong Liu and Zekun Wang and Zeyu Cui and Zhenru Zhang and Zhipeng Zhou and Zihan Qiu},
      year={2025},
      eprint={2505.09388},
      archivePrefix={arXiv},
      primaryClass={cs.CL},
      url={https://arxiv.org/abs/2505.09388}, 
}

@misc{jiang2023mistral7b,
      title={Mistral 7B}, 
      author={Albert Q. Jiang and Alexandre Sablayrolles and Arthur Mensch and Chris Bamford and Devendra Singh Chaplot and Diego de las Casas and Florian Bressand and Gianna Lengyel and Guillaume Lample and Lucile Saulnier and Lélio Renard Lavaud and Marie-Anne Lachaux and Pierre Stock and Teven Le Scao and Thibaut Lavril and Thomas Wang and Timothée Lacroix and William El Sayed},
      year={2023},
      eprint={2310.06825},
      archivePrefix={arXiv},
      primaryClass={cs.CL},
      url={https://arxiv.org/abs/2310.06825}, 
}

@misc{deepseekai2024deepseekllmscalingopensource,
      title={DeepSeek LLM: Scaling Open-Source Language Models with Longtermism}, 
      author={DeepSeek-AI and Xiao Bi and Deli Chen and Guanting Chen and Shanhuang Chen and Damai Dai and Chengqi Deng and Honghui Ding and Kai Dong and Qiushi Du and Zhe Fu and Huazuo Gao and Kaige Gao and Wenjun Gao and Ruiqi Ge and Kang Guan and Daya Guo and Jianzhong Guo and Guangbo Hao and Zhewen Hao and Ying He and Wenjie Hu and Panpan Huang and Erhang Li and Guowei Li and Jiashi Li and Yao Li and Y. K. Li and Wenfeng Liang and Fangyun Lin and A. X. Liu and Bo Liu and Wen Liu and Xiaodong Liu and Xin Liu and Yiyuan Liu and Haoyu Lu and Shanghao Lu and Fuli Luo and Shirong Ma and Xiaotao Nie and Tian Pei and Yishi Piao and Junjie Qiu and Hui Qu and Tongzheng Ren and Zehui Ren and Chong Ruan and Zhangli Sha and Zhihong Shao and Junxiao Song and Xuecheng Su and Jingxiang Sun and Yaofeng Sun and Minghui Tang and Bingxuan Wang and Peiyi Wang and Shiyu Wang and Yaohui Wang and Yongji Wang and Tong Wu and Y. Wu and Xin Xie and Zhenda Xie and Ziwei Xie and Yiliang Xiong and Hanwei Xu and R. X. Xu and Yanhong Xu and Dejian Yang and Yuxiang You and Shuiping Yu and Xingkai Yu and B. Zhang and Haowei Zhang and Lecong Zhang and Liyue Zhang and Mingchuan Zhang and Minghua Zhang and Wentao Zhang and Yichao Zhang and Chenggang Zhao and Yao Zhao and Shangyan Zhou and Shunfeng Zhou and Qihao Zhu and Yuheng Zou},
      year={2024},
      eprint={2401.02954},
      archivePrefix={arXiv},
      primaryClass={cs.CL},
      url={https://arxiv.org/abs/2401.02954}, 
}

@inproceedings{
hu2022lora,
title={Lo{RA}: Low-Rank Adaptation of Large Language Models},
author={Edward J Hu and yelong shen and Phillip Wallis and Zeyuan Allen-Zhu and Yuanzhi Li and Shean Wang and Lu Wang and Weizhu Chen},
booktitle={International Conference on Learning Representations},
year={2022},
url={https://openreview.net/forum?id=nZeVKeeFYf9}
}

@inproceedings{zheng-etal-2024-helpful,
    title = "When ``A Helpful Assistant'' Is Not Really Helpful: Personas in System Prompts Do Not Improve Performances of Large Language Models",
    author = "Zheng, Mingqian  and
      Pei, Jiaxin  and
      Logeswaran, Lajanugen  and
      Lee, Moontae  and
      Jurgens, David",
    editor = "Al-Onaizan, Yaser  and
      Bansal, Mohit  and
      Chen, Yun-Nung",
    booktitle = "Findings of the Association for Computational Linguistics: EMNLP 2024",
    month = nov,
    year = "2024",
    address = "Miami, Florida, USA",
    publisher = "Association for Computational Linguistics",
    url = "https://aclanthology.org/2024.findings-emnlp.888/",
    doi = "10.18653/v1/2024.findings-emnlp.888",
    pages = "15126--15154",
}

@misc{chatalpaca2023,
  author = {Ning Bian and Hongyu Lin and Yaojie Lu and Xianpei Han and Le Sun and Ben He },
  title = {ChatAlpaca: A Multi-Turn Dialogue Corpus based on Alpaca Instructions},
  year = {2023},
  publisher = {GitHub},
  journal = {GitHub repository},
  howpublished = {\url{https://github.com/cascip/ChatAlpaca}},
}

@misc{sharegpt2023,
  title        = {ShareGPT GPT-4 Dataset},
  author       = {Shibing624},
  year         = {2023},
  howpublished = {\url{https://huggingface.co/datasets/shibing624/sharegpt_gpt4}},
}

@inproceedings {bagdasaryan2021blind,
author = {Eugene Bagdasaryan and Vitaly Shmatikov},
title = {Blind Backdoors in Deep Learning Models},
booktitle = {30th USENIX Security Symposium (USENIX Security 21)},
year = {2021},
isbn = {978-1-939133-24-3},
pages = {1505--1521},
url = {https://www.usenix.org/conference/usenixsecurity21/presentation/bagdasaryan},
publisher = {USENIX Association},
month = aug
}

@inproceedings {guo2025persistent,
author = {Zhen Guo and Abhinav Kumar and Reza Tourani},
title = {Persistent Backdoor Attacks in Continual Learning},
booktitle = {34th USENIX Security Symposium (USENIX Security 25)},
year = {2025},
isbn = {978-1-939133-52-6},
address = {Seattle, WA},
pages = {6379--6397},
url = {https://www.usenix.org/conference/usenixsecurity25/presentation/guo-zhen},
publisher = {USENIX Association},
month = aug
}


\end{document}